\documentclass[a4paper,11pt]{article}
\makeatletter
\@namedef{ver@natbib.sty}{9999/99/99}
\@namedef{opt@natbib.sty}{numbers,sort&compress}
\makeatother

\usepackage{pos}
\makeatletter
\expandafter\let\csname ver@natbib.sty\endcsname\@undefined
\makeatother
\usepackage[backend=biber,style=numeric-comp,url=true,sorting=none,giveninits=true,
            maxbibnames=99,doi=false,isbn=false,eprint=true]{biblatex}
\DeclareNameAlias{default}{given-family}

\DeclareFieldFormat*{title}{\mkbibemph{#1}}
\DeclareFieldFormat{booktitle}{proceedings of \mkbibemph{#1}}
\DeclareFieldFormat{eprint:arxiv}{[\texttt{arXiv:#1}]}
\renewcommand*{\intitlepunct}{\space}
\renewbibmacro*{in:}{\ifentrytype{article}{}{\printtext{\bibstring{in}\intitlepunct}}}
\AtEveryBibitem{\ifentrytype{article}{\clearname{editor}\clearfield{number}}{}}
\DeclareFieldFormat[article]{pages}{#1}

\DeclareFieldFormat{url}{\url{#1}}
\renewbibmacro*{eprint}{%
  \setunit{\addspace}%
  \iffieldundef{eprinttype}
    {\printfield{eprint}}
    {\printfield[eprint:\strfield{eprinttype}]{eprint}}}
\DeclareFieldFormat{number}{#1}
\AtEveryBibitem{\clearfield{institution}\clearlist{institution}}

\title{Machine learning for trigger, reconstruction and generation}
\ShortTitle{ML for trigger, reconstruction and generation}

\author*{Marco Giacalone}

\affiliation{European Organization for Nuclear Research (CERN),\\
  Geneva, Switzerland}

\emailAdd{mgiacalo@cern.ch}

\abstract{Machine learning (ML) is increasingly embedded in every stage of the LHC
data-processing chain, from event generation to final reconstruction. This
contribution reviews the latest ML developments from the ALICE, ATLAS, CMS and
LHCb collaborations that were made public in the year preceding
LHCP2026, following the workflow of the experiments: event generation
and detector simulation, real-time trigger selection, and online/offline
reconstruction. Highlights include generative models that accelerate the
simulation of specific ALICE detector components by up to six orders of
magnitude, the CMS AXOL1TL and ATLAS GELATO unsupervised anomaly-detection
triggers now operating in Run~3, a quantised Deep Sets jet tagger for the
CMS Phase-2 Level-1 trigger, ML-based jet energy correction and flavour
tagging at LHCb, the first empirical scaling laws for transformer-based
flavour taggers in ATLAS, and several transformer and graph-based
reconstruction algorithms (ATLAS GN3 flavour tagging and MaskFormer tracking,
LHCb GarNet,
CMS MLPF) developed for HL-LHC operation.}

\FullConference{14th Edition of the Large Hadron Collider Physics (LHCP2026)\\
  18-22 May 2026\\
  Paris, France\\}

\begin{document}
\maketitle

\section{Introduction}
\label{sec:intro}
In a typical experiment, collision data are recorded by the detector, selected online by a trigger system, and processed by a reconstruction chain to produce physics objects. In parallel, Monte Carlo events undergo an analogous processing chain, from event generation and detector transport to reconstruction. Machine learning (ML) techniques can be inserted at any point of this workflow to
improve performance, precision, or computational efficiency, and detector simulation can be
regarded as the opposite path starting from simulated particles, then tracks and finally detector signals. This contribution
reviews such developments across the ALICE, ATLAS, CMS and LHCb
collaborations, focusing on results made public since LHCP2025. 
Given the extent of the subject, the selection is
illustrative rather than exhaustive, and applications of ML to physics
analyses are not discussed. The material follows the data workflow itself:
Sect.~\ref{sec:sim} covers event generation and detector simulation,
Sect.~\ref{sec:trigger} the trigger, Sect.~\ref{sec:reco} reconstruction, and
Sect.~\ref{sec:summary} provides a summary and outlook.
\section{Simulation: accelerating event generation and detector response}
\label{sec:sim}
Monte Carlo production is one of the most computationally expensive
activities of an LHC experiment, and ML-based surrogate models are being
adopted at every step of it. Replacing event generation alone already removes
a significant bottleneck, but full detector transport, typically performed
with Geant4, remains the dominant cost; replacing transport as well
moves the bottleneck downstream, to the trigger and reconstruction steps
discussed in Sects.~\ref{sec:trigger} and~\ref{sec:reco}. Calorimeters are a
particularly well-known bottleneck, because of the many secondary particles
produced in showers, and have motivated substantial community effort,
exemplified by the CaloChallenge 2022 initiative~\cite{Krause:2024avx}. 
A more radical approach, known as ``flash'' simulation, skips the entire
simulation chain and generates analysis-level objects directly from generator-level
truth. CMS and LHCb have both reported earlier progress on this, but no new public results appeared in the period covered here.
\subsection{Fast simulation of TPC loopers with a WGAN-GP}
\label{sec:loopers}
\begin{figure}[!htbp]
\centering
\includegraphics[width=0.545\textwidth]{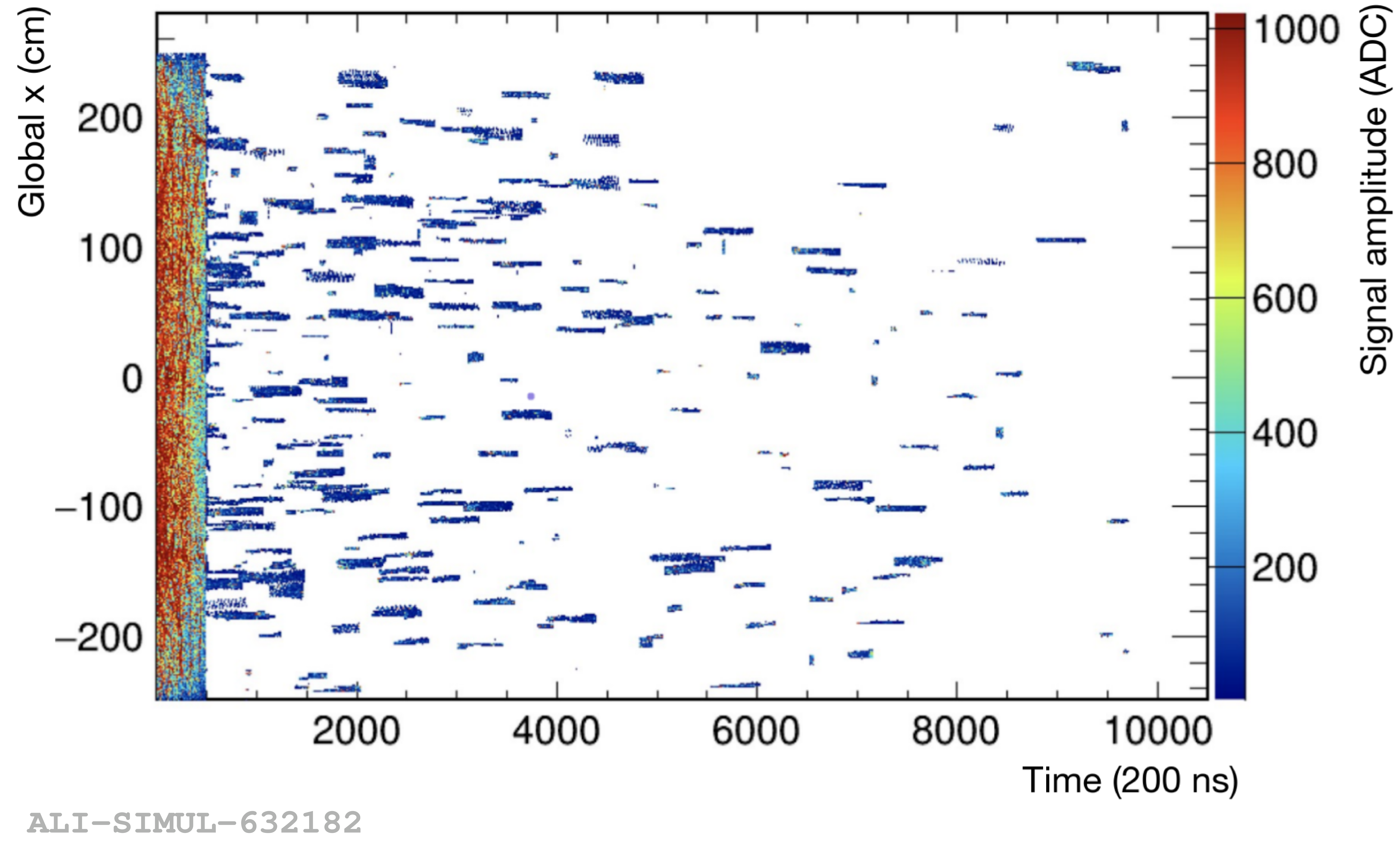}\hfill
\includegraphics[width=0.415\textwidth]{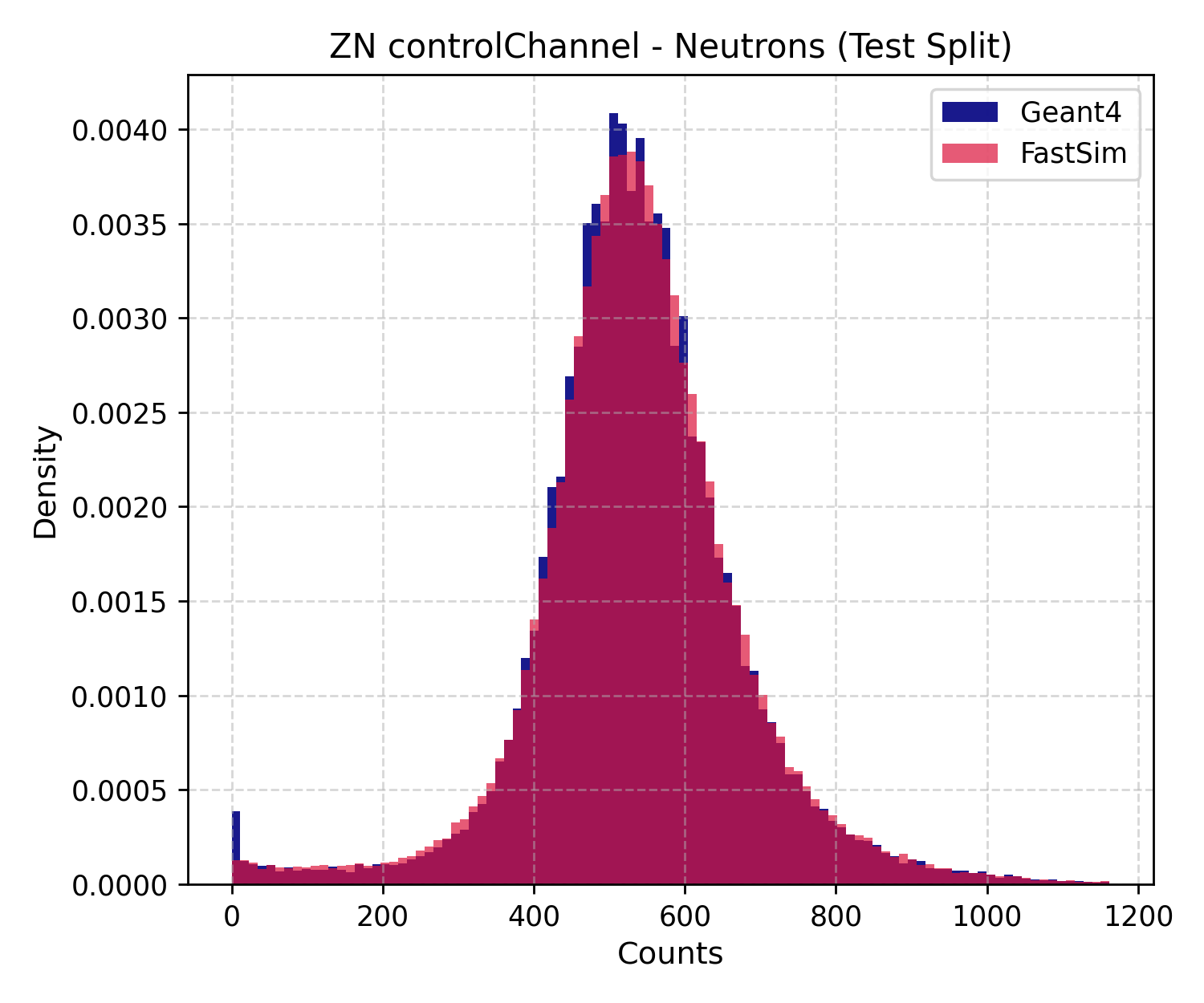}
\caption{(Left) TPC looper
clusters, shown as global $x$ position as a function of time, produced by the
trained WGAN-GP model and injected into a minimum-bias event. (Right)
Energy-deposit spectrum for neutrons in the ZN calorimeter, comparing the
Geant4 reference (blue) with the CFM fast simulation (red).}
\label{fig:alicesim}
\end{figure}
Slow neutrons interacting with the enclosure of the ALICE Time Projection Chamber
(TPC) indirectly produce low-momentum e$^+$e$^-$ pairs and Compton electrons that spiral repeatedly inside the
detector volume, leaving long-lived, densely populated tracks known as
``loopers''. Transporting the slow neutrons through the full Geant4 chain is very
expensive and makes their inclusion in standard productions impractical on
standard Grid resources. ALICE therefore trained a Wasserstein generative
adversarial network with gradient penalty (WGAN-GP) to reproduce the final electrons and positrons 
from slow-neutron interactions, that are later injected in the transport step: a generator maps
noise to the physical features of electrons and positrons, while a critic scores real and generated particles
so as to approximate the Wasserstein distance between the two distributions.
The trained generator injects synthetic loopers into minimum-bias events at the transport stage,
reproducing the cluster occupancy observed in real events with a factor 10
speed-up; the generated clusters are shown in Fig.~\ref{fig:alicesim} (left).
The method is already used in production.
\subsection{Conditional flow matching for the Zero Degree Calorimeter}
\label{sec:zdc}
The ALICE Zero Degree Calorimeter (ZDC) sits far downstream of the
interaction point, and its inclusion significantly increases the running time of
a standard simulation workflow; it is therefore excluded from default Monte
Carlo productions. ALICE is studying Conditional Flow Matching (CFM) as a remedy:
the model learns a time-dependent velocity field that transports a Gaussian
source distribution into the target distributions of energy
deposits. It reproduces the Geant4 reference distributions in both
the neutron (ZN in Fig.~\ref{fig:alicesim}, right) and proton calorimeter control
channels while running about six orders of magnitude faster than full
transport, making it realistic to include the ZDC routinely in future
productions.

\section{Trigger: real-time selection with machine learning}
\label{sec:trigger}
ML is being applied to both the hardware and software components of LHC trigger systems,
to handle higher data rates while maintaining physics performance and to increase selection efficiency.
It also aims to detect anomalous event topologies that conventional, model-dependent triggers might miss,
while satisfying the severe constraints defined by the detectors.
\subsection{Unsupervised anomaly detection in CMS: AXOL1TL and CICADA}
\label{sec:axol1tl}
\begin{figure}[htbp]
\centering
\includegraphics[width=0.65\textwidth]{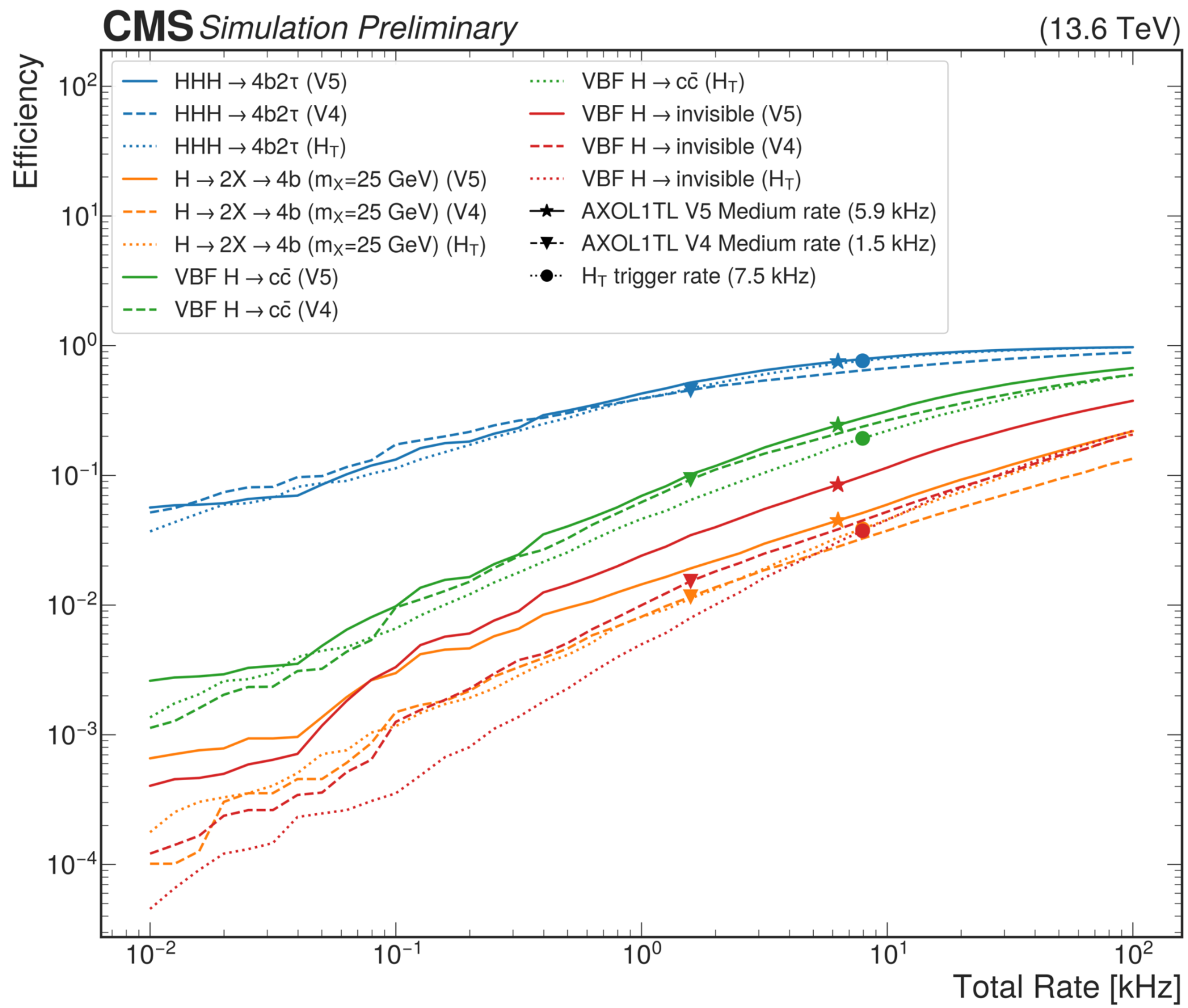}
\caption{Efficiency of AXOL1TL as a function of
the total trigger rate, for several beyond-Standard-Model benchmark signals,
comparing version~5 (solid) and version~4 (dashed) with an $H_{\mathrm{T}}$-based
trigger (dotted). Markers indicate the medium-rate working points of the two
AXOL1TL versions and the $H_{\mathrm{T}}$ trigger~\cite{cms:axol1tl}.}
\label{fig:axol1tl}
\end{figure}

CMS operates two complementary unsupervised anomaly-detection algorithms in
its Level-1 hardware trigger during Run~3: AXOL1TL, based on global
trigger-object inputs, and CICADA, based on Layer-1 calorimeter-tower inputs.
Both are trained exclusively on Standard Model data, so that a large anomaly
score signals a departure from the trained distribution regardless of the
underlying physics process. They are therefore signal-agnostic by
construction and extend, rather than replace, the dedicated supervised
triggers used for specific final states. In 2025, AXOL1TL was updated to its
fifth version, which adds a Variance-Invariance-Covariance Regularisation
(VICReg) feature extractor on top of the existing variational autoencoder and
redefines the anomaly score as the squared reconstruction error,
$\sum_i (x_i-x_i^\prime)^2$, rather than the mean encoder output used
previously. Figure~\ref{fig:axol1tl} compares the efficiency of the fourth
and fifth versions, at their respective medium-rate working points, with a
conventional trigger based on the scalar sum of jet transverse momenta
($H_{\mathrm{T}}$), for a range of benchmark signals. At equal total rate the fifth
version is more efficient than the fourth for most of the benchmarks
considered, and at its working point it reaches a higher efficiency than the
$H_{\mathrm{T}}$ trigger while running at a lower rate; a sixth version is under
development.
The full pipeline is converted for FPGA deployment with \texttt{hls4ml} and
operates within the strict latency budget of the Level-1 trigger.
\subsection{GELATO: the first ATLAS anomaly-detection trigger}
\label{sec:gelato}
ATLAS has deployed its first anomaly-detection trigger, GELATO (a Generic
Event-Level Anomalous Trigger Option)~\cite{atlas:gelato}, integrated at both
Level~1, using a variational autoencoder combined with a generative
adversarial network (VAE-GAN), and at the High Level Trigger (HLT), using a
simpler autoencoder running on events already selected by the GELATO L1
trigger. In the VAE-GAN, a discriminator sharpens the reconstruction with respect to a
standard VAE, and the training loss combines a masked mean-squared-error term,
a Kullback-Leibler (KL) divergence term and the adversarial loss; only the encoder
is retained for inference, given the limited FPGA resources available online.
Both models were trained on a 2024 enhanced-bias dataset with detector
conditions close to those of 2025. The trigger currently runs at reduced rates of
1~Hz at L1 and 0.1~Hz at the HLT, with studies ongoing to reach the target
rates of 500~Hz and 10~Hz.

\subsection{Deep Sets jet tagging for the CMS Phase-2 Level-1 trigger}
\label{sec:deepsets}
Looking ahead to the High-Luminosity LHC, CMS is developing a multiclass jet
tagger for the Phase-2 upgrade of the Level-1 trigger, designed to provide
high-priority selection of \mbox{(di-)}Higgs production while coping with the
increased luminosity~\cite{cms:deepsets}. The tracking information added by
the Phase-2 upgrade makes it possible to move beyond purely kinematic triggers
and to perform constituent-level analysis directly in hardware. The chosen
architecture is a quantised Deep Sets network, whose permutation invariance suits an FPGA
implementation: up to 16 $p_{\mathrm{T}}$-sorted particle-flow candidates per jet, each
with 20 input features, are processed by convolutional layers and
combined by global-average pooling, before branching into a classification
head providing eight jet-flavour categories (light, charm, $b$, gluon,
hadronic taus of either charge, electrons and muons) and a regression head for
the jet $p_{\mathrm{T}}$. The implementation fits the hard constraints of the Level-1
trigger (L1T) FPGA and improves the selection efficiency for the $HH\to4b$
channel by roughly 15\% as a function of the total jet-$p_{\mathrm{T}}$ sum.
\begin{figure}[!ht]
\centering
\includegraphics[width=0.45\textwidth]{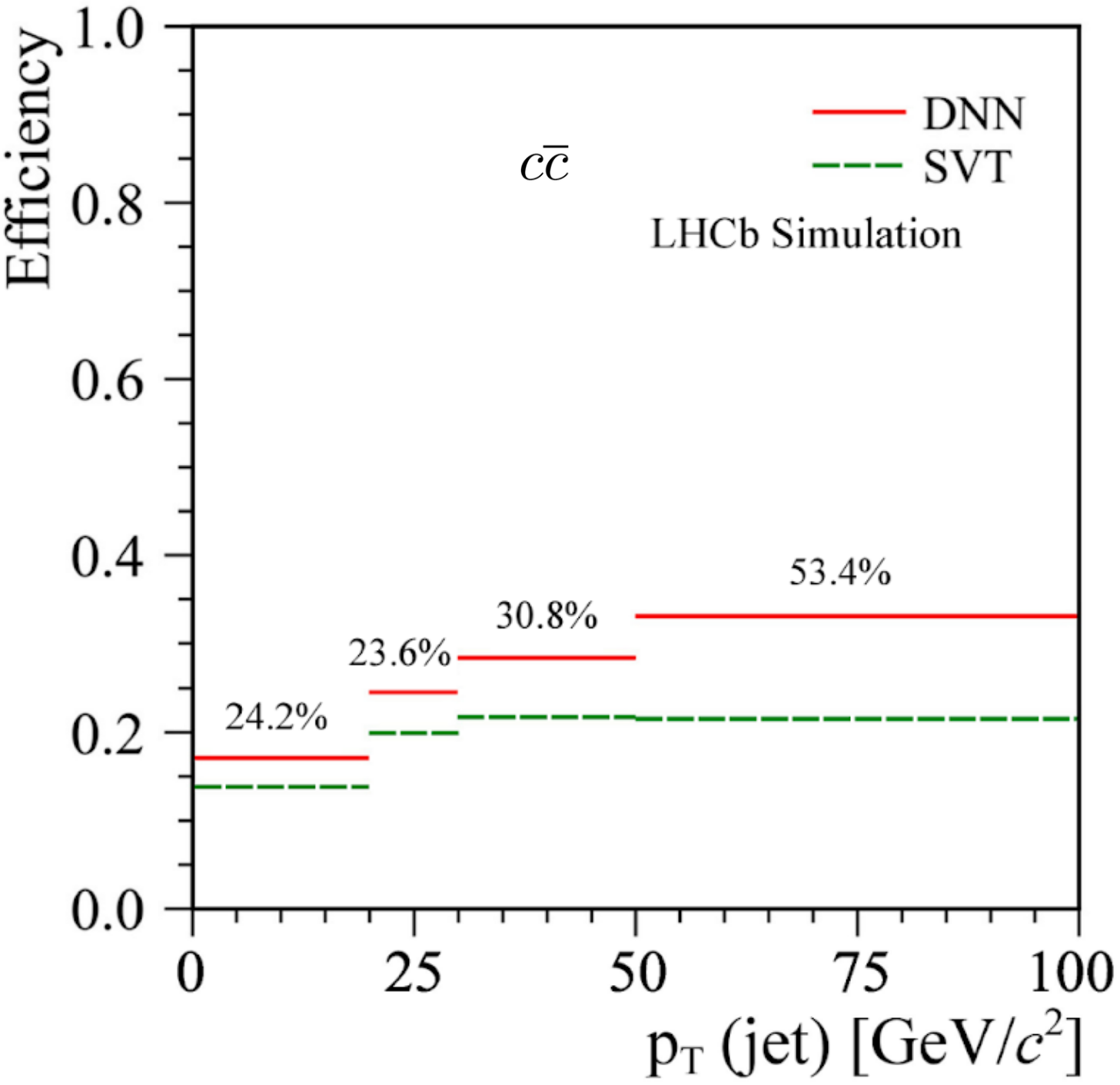}
\caption{Tagging efficiency of the new Deep Neural Network (DNN) classifier compared to the
standard secondary-vertex tagger (SVT) as a function of jet $p_{\mathrm{T}}$, for jets
originating from Higgs-boson decays to $c\bar c$. Percentages indicate the relative improvement in each
interval~\cite{lhcb:jettag}.}
\label{fig:lhcbjet}
\end{figure}
\section{Reconstruction: machine learning across the offline and online chain}
\label{sec:reco}

Event reconstruction comprises many distinct components (track finding,
clustering, vertexing, jet building, flavour tagging), each an independent
target for optimisation with ML, and one may even ask whether an entirely
ML-based reconstruction could eventually replace the traditional sequential
chain. The following examples illustrate progress over the past year.

\subsection{Jet reconstruction and flavour tagging at LHCb}
\label{sec:lhcbjet}
LHCb introduced two new ML-based methods to improve jet reconstruction,
motivated by the search for $H\to b\bar b$ and $H\to c\bar c$
decays~\cite{lhcb:jettag}. Because $b$- and $c$-jets tend to be reconstructed
with a larger energy underestimate than light jets, a dedicated
regression-based jet-energy correction, using a Gradient Boosting Regressor,
is applied to improve the dijet invariant-mass resolution of Higgs-boson
candidates. Flavour tagging is instead performed with a deep neural network,
inspired by the CMS DeepJet architecture, combining a one-dimensional
convolutional layer acting on per-particle features, a recursive (long
short-term memory) layer, and a dense output layer providing $b$/$c$/light-quark
probabilities. The new tagger improves the single-jet tagging efficiency with
respect to the standard secondary-vertex tagging algorithm (SVT) by
9.8--27.3\% for $b$-jets and by 23.6--53.4\% for $c$-jets, depending on the
jet $p_{\mathrm{T}}$ interval (the $c\bar c$ case is shown in Fig.~\ref{fig:lhcbjet}). Applied to the 2016 dataset
(1.6~fb$^{-1}$), no significant signal is observed, but upper limits are set
on the inclusive production cross section (Table~\ref{tab:lhcblimits});
extrapolations indicate that an observation of $H\to b\bar b$ may become
feasible with the dataset collected by the end of Run~4.
\begin{table}[htbp]
\centering
\footnotesize
\begin{tabular}{l|ccc|ccc}
\hline
 & \multicolumn{3}{c|}{Expected limits} & \multicolumn{3}{c}{Observed limits}\\
Process & $\sigma_{\rm UP}/\sigma_{\rm SM}$ & $N_{\rm UP}$ & $\sigma_{\rm UP}$~[pb]
        & $\sigma_{\rm UP}/\sigma_{\rm SM}$ & $N_{\rm UP}$ & $\sigma_{\rm UP}$~[pb]\\
\hline
$H\to b\bar b$ & 11.1 & 2652 & 353  & 6.64 & 1587 & 211 \\
$H\to c\bar c$ & 1834 & 1414 & 2939 & 1003 & 722  & 1605 \\
\hline
\end{tabular}
\caption{Expected and observed upper limits on the
inclusive production of $H\to b\bar b$ and $H\to c\bar c$, obtained with the
ML-based jet reconstruction on the 2016 LHCb dataset~\cite{lhcb:jettag}.}
\label{tab:lhcblimits}
\end{table}

\subsection{A neural-network cluster finder for the ALICE TPC}
\label{sec:clusterizer}
The online cluster finder for the ALICE TPC is currently a GPU-based
algorithm; a natural extension, targeting Run~4 and beyond, is to augment it
with two neural networks: a classification network that discards clusters
unlikely to originate from a genuine track, and a regression network that
refines the cluster properties used for particle identification. The former
reduces the number of stored clusters by up to 18\% while leaving the
reconstructed $\mathrm{K^0_S\to\pi^+\pi^-}$, $\Lambda\to\pi p$ and $\mathrm{D^0\to K^-\pi^+}$
(+ c.c.) invariant-mass peaks unaffected, and it saves disk space by a comparable percentage; 
the latter improves the $\mathrm{d}E/\mathrm{d}x$ separation between electrons and pions. Validated on
commissioning pp collisions at 500~kHz, this was the first application of
neural networks in ALICE online processing, and two additional Pb--Pb
commissioning runs have been successfully performed in the beginning of June.

\subsection{Scaling laws for jet-flavour tagging in ATLAS}
\label{sec:carpedatum}
\begin{figure}[htbp]
\centering
\includegraphics[width=0.7\textwidth]{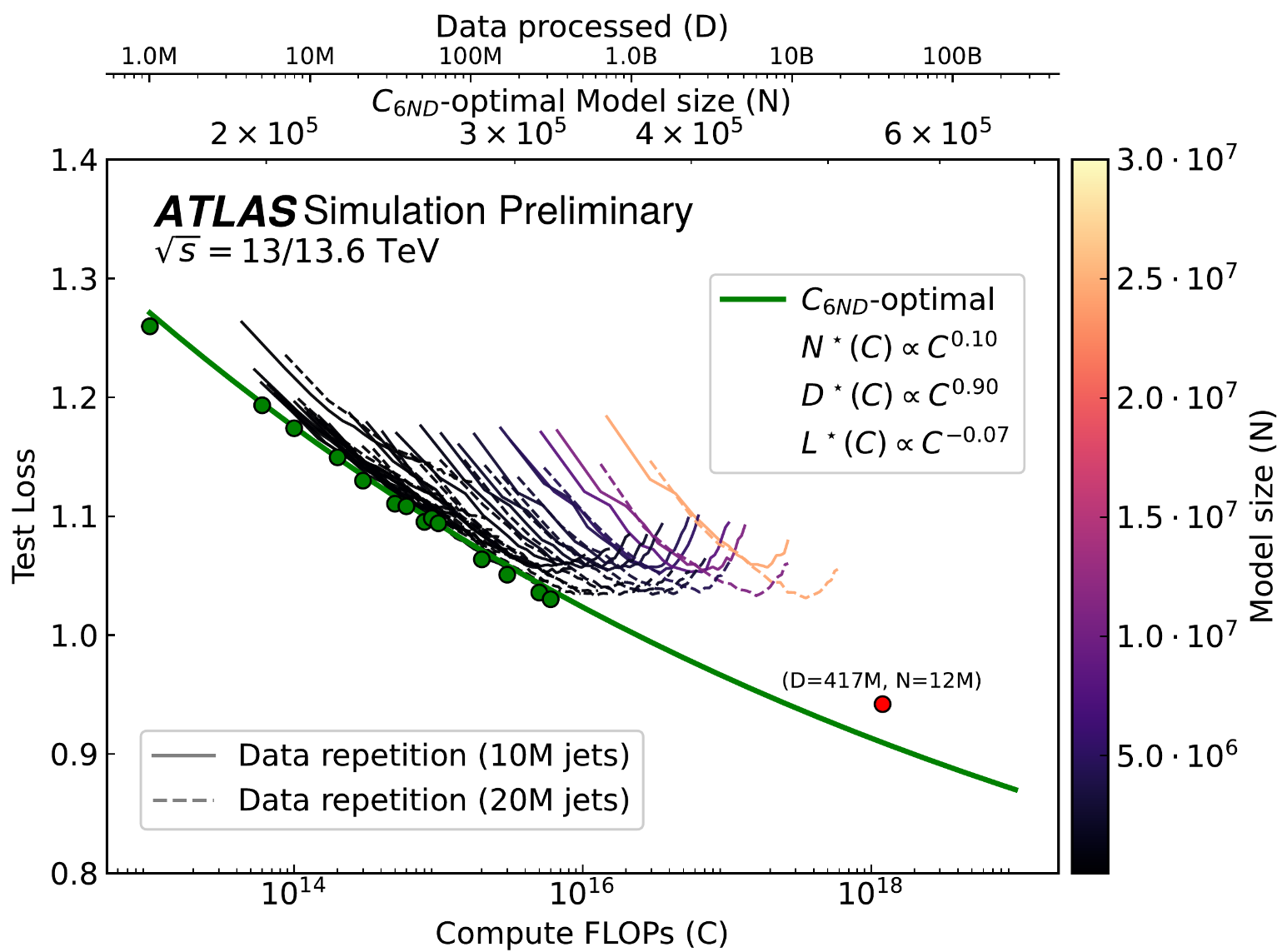}
\caption{Test loss versus training compute for transformer-based flavour
taggers of varying size, with the compute-optimal trajectory (solid line)
and the fitted scaling exponents indicated; the isolated marker shows the
current 12-million-parameter model trained on 417~million
jets~\cite{atlas:carpedatum}.}
\label{fig:carpedatum}
\end{figure}

Motivated by the empirical scaling laws describing the performance of
large language models as a function of model size, dataset size and compute,
ATLAS investigated whether transformer-based flavour taggers such as GN2
follow a similar behaviour~\cite{atlas:carpedatum}. The usual compute cost,
$C=6ND$, with $N$ the number of trainable parameters and $D$ the number of
training jets, was extended with two terms specific to high-energy physics,
accounting for the per-sample simulation cost and for repeated inference
across multiple analyses; the latter penalises over-parametrised models and
pushes the compute-optimal trajectory towards smaller architectures. Fitting
the loss surface $L(N,D)$ over a grid of model and dataset sizes, in a
single-epoch regime, yields a smooth power-law behaviour
with $N^\star(C)\propto C^{0.10}$, $D^\star(C)\propto C^{0.90}$ and
$L^\star(C)\propto C^{-0.07}$ (Fig.~\ref{fig:carpedatum}). 
The current-generation 12-million-parameter tagger, trained on
417~million jets, was found to sit off this compute-optimal curve. Guided by
the extracted scaling law, an 86-million-parameter tagger was subsequently
trained on 7.7~billion jets, improving the background rejection by up to a
factor 1.8 for $c$-jets and 1.4 for light jets at low $b$-jet efficiency
working points, consistent with the empirical prediction and confirming that
current flavour taggers are still far from saturating their achievable
performance.

\subsection{Transformers for tracking in dense environments}
\label{sec:maskformer}
Standard combinatorial tracking algorithms in ATLAS struggle inside the dense
cores of high-$p_{\mathrm{T}}$ jets, where multiple particles can contribute to a
single pixel cluster and the usual one-cluster-per-track assumption breaks
down. As a proof of concept, ATLAS proposed a transformer-based
encoder-decoder model built on the MaskFormer architecture, which treats
track reconstruction as an instance-segmentation problem: learnable
object-level queries are trained to represent individual charged particles
and to assign hits to them through attention, which naturally supports
assigning a shared cluster to several tracks rather than discarding it, as
the standard reconstruction does~\cite{atlas:maskformer}. In a hadronic
region of interest, the model improves the charged-particle reconstruction
efficiency by up to 30\% at high $p_{\mathrm{T}}$ with respect to both the baseline
tracks and the reference silicon space-point seeded tracks: in
$Z^\prime\to q\bar q$ events the efficiency stays close to 0.8 above
1~TeV, where the standard reconstruction falls below 0.6. Cluster confusion
matrices confirm that shared clusters are correctly assigned to multiple
tracks. Further studies are foreseen before the method can be
considered for production use.

\subsection{GN3: the next-generation ATLAS jet-flavour tagger}
\label{sec:gn3}
\begin{figure}[htbp]
\centering
\includegraphics[width=0.65\textwidth]{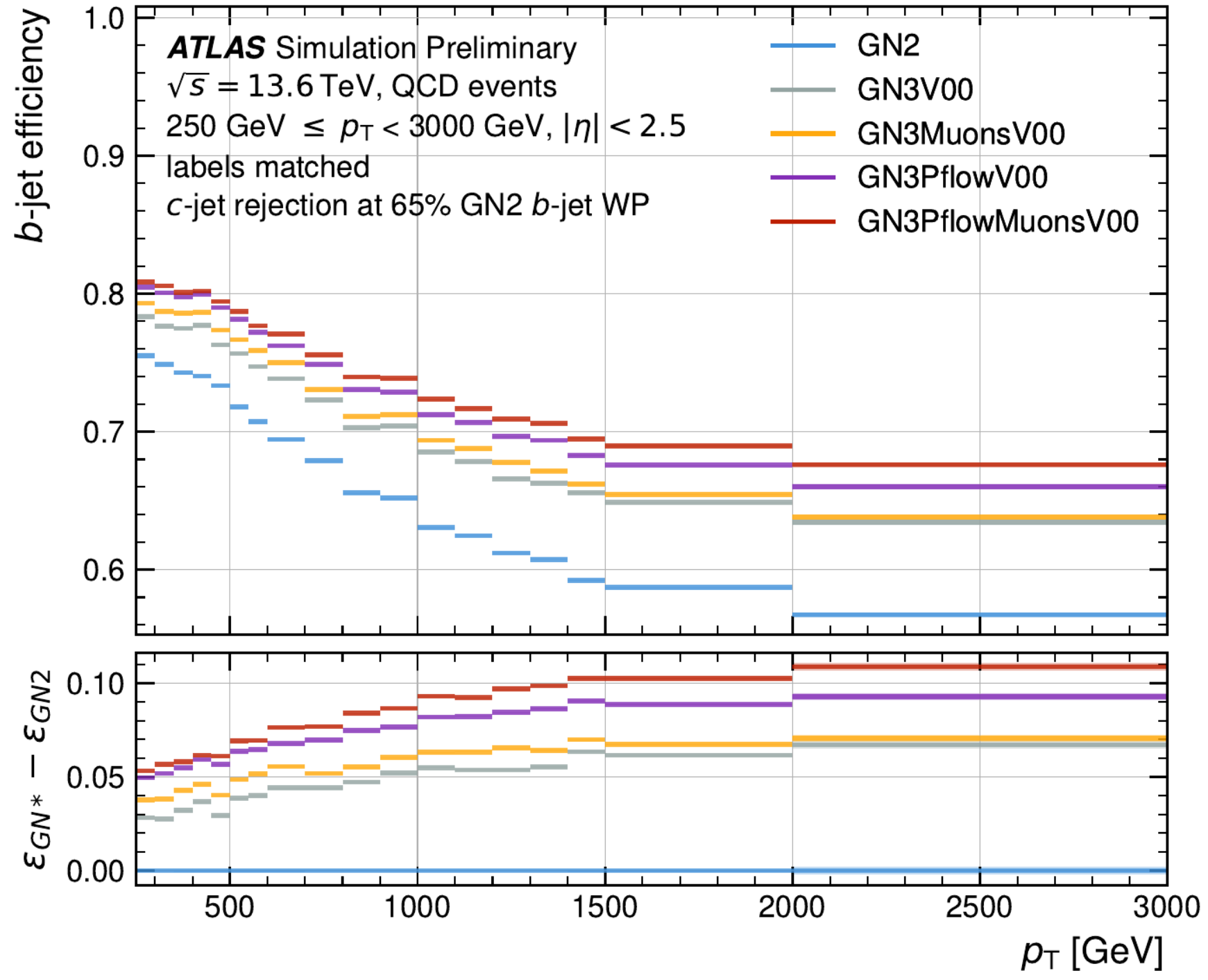}
\caption{$b$-jet efficiency versus jet $p_{\mathrm{T}}$ in QCD events at a fixed GN2
working point, comparing GN2 with several GN3 input configurations; the lower
panel shows the absolute efficiency gain with respect to
GN2~\cite{atlas:gn3}.}
\label{fig:gn3}
\end{figure}

Building on GN2, ATLAS has developed GN3, a next-generation flavour-tagging
algorithm that adopts a multi-modal input representation, jointly processing
charged and neutral particle-flow objects together with an extended set of
track features and looser selection cuts, so as to exploit more fully the
capacity of its transformer-based architecture~\cite{atlas:gn3}. The
four-layer transformer core is retained from GN2, but several components are
upgraded: flash attention improves memory efficiency for variable-length
input sequences, gated linear units improve gradient flow during training,
and an attention-pooling stage summarises the encoder output into a global
jet representation, providing a richer context than the track-only summary
used in GN2. The gains are substantial: in $t\bar t$ events, GN3 improves the
background rejection with respect to GN2 by roughly a factor 1.8 for
$c$-jets, 2.3 for light jets and up to 5 for $\tau$-jets, while in
high-$p_{\mathrm{T}}$ Quantum Chromodynamics (QCD) events the $b$-jet efficiency at a fixed GN2 working point
increases by up to 0.10 in absolute terms (Fig.~\ref{fig:gn3}).

\subsection{GarNet: a lightweight graph neural network for LHCb}
\label{sec:garnet}
For the LHCb Upgrade~II, operating at five times the current instantaneous
luminosity and data rates of about 200~Tb/s, particle identification in the
future PicoCal electromagnetic calorimeter, which will also provide
$\mathcal{O}(10)$~ps timing information, must be both fast and accurate.
GarNet was developed as a lightweight, attention-based graph neural network
alternative to traditional message-passing GNNs for this task: two
attention-based layers with four learnable aggregators each let the model
capture the global structure of a cluster, while removing the explicit edge
features between hits further reduces the computational cost. The resulting
model is approximately eight times faster than the message-passing baseline
and achieves a better energy resolution over the full range of incoming
transverse energy ($E_{\mathrm{T}}$), especially at low values, where the resolution for single
photons with minimum-bias overlay improves from about 0.27 to
0.16~\cite{lhcb:garnet}.

\subsection{Towards an end-to-end machine-learning particle flow in CMS}
\label{sec:mlpf}
\begin{figure}[htbp]
\centering
\includegraphics[width=0.45\textwidth]{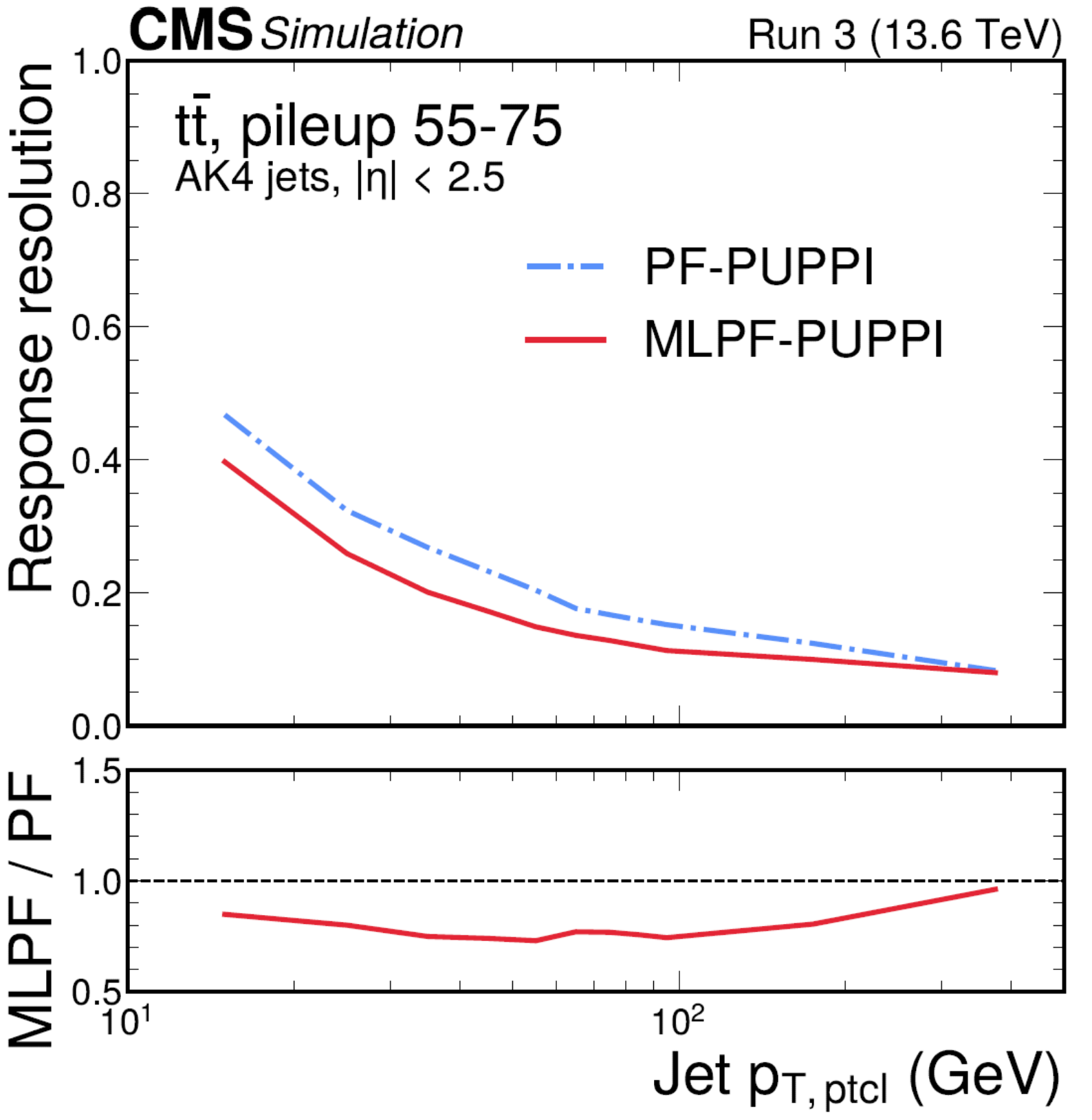}\hfill
\includegraphics[width=0.45\textwidth]{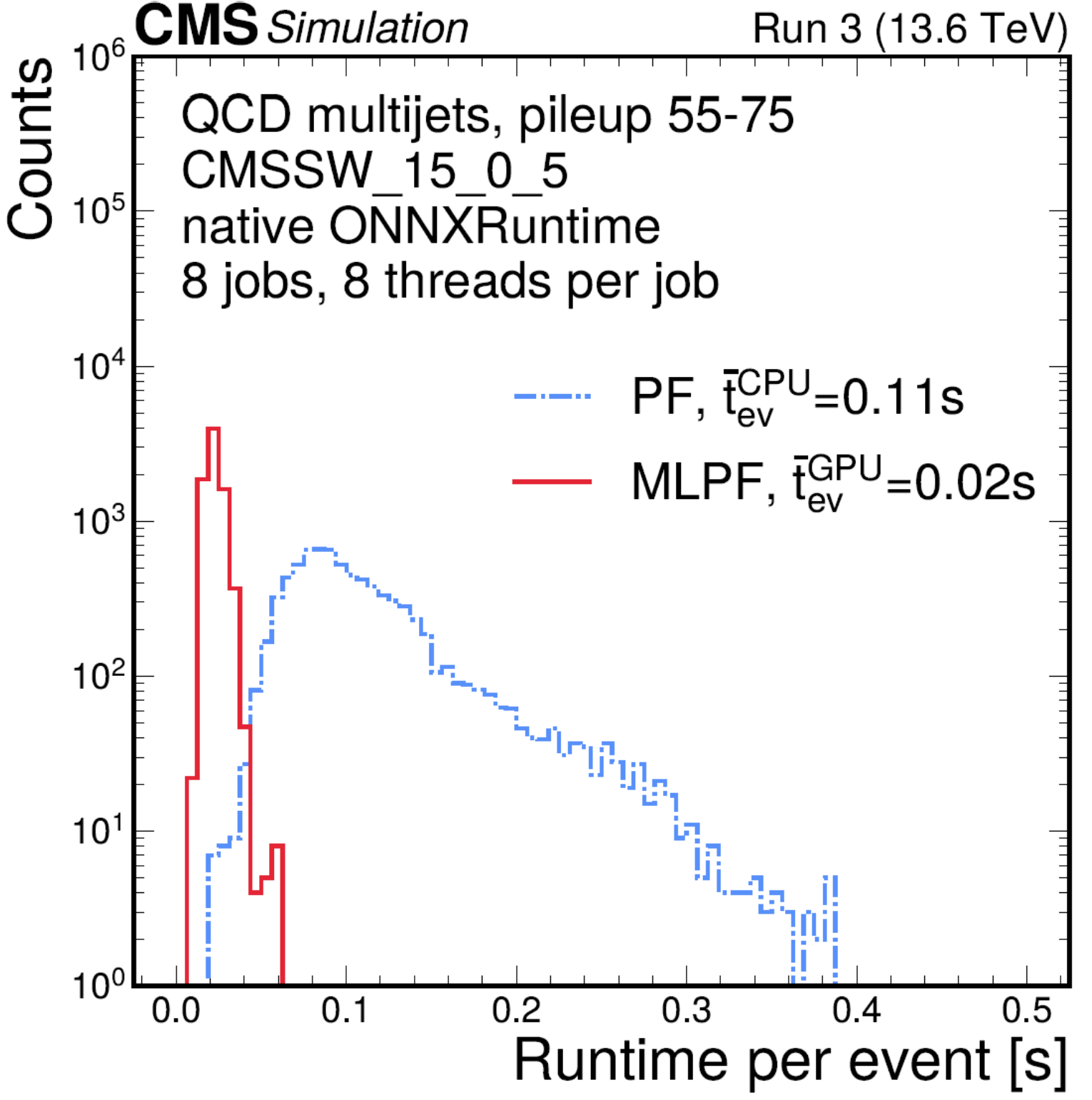}
\caption{(Left) Jet response resolution as a function of the particle-level
jet $p_{\mathrm{T}}$ in $t\bar t$ events, comparing MLPF with the standard
particle-flow algorithm. (Right) Distribution of the processing time per
event for QCD multijet events~\cite{cms:mlpf}.}
\label{fig:mlpf}
\end{figure}

The particle-flow (PF) algorithm, which combines tracking and calorimetry
into a single list of reconstructed particles, has been the basis of CMS
event reconstruction since Run~1. Its reliance on hard-coded linking rules
between tracks and calorimeter clusters, however, leads to rigid decision
boundaries, requires substantial re-tuning whenever the detector geometry
changes, and scales poorly with event complexity. CMS is therefore exploring
a machine-learning-based particle-flow algorithm (MLPF) that replaces this
logic with a learned, data-driven approach, implemented as a transformer with
self-attention that predicts all reconstructed particles of an event in a
single inference step, natively on GPU~\cite{cms:mlpf}. The model was
trained, validated and tested on simulated Run~3 pp collisions and achieves
physics performance comparable to the standard algorithm; the jet response
resolution in $t\bar t$ events is in fact up to about 25\% better than
that of the standard PF algorithm in the intermediate $p_{\mathrm{T}}$ range, converging
with it at the highest $p_{\mathrm{T}}$ (Fig.~\ref{fig:mlpf}, left), despite the
model being optimised at the level of individual particles rather than jets.
At the same time, the average processing time per event drops from 0.11~s on
CPU to 0.02~s on GPU (Fig.~\ref{fig:mlpf}, right).

\section{Summary and outlook}
\label{sec:summary}
In a single year from LHCP2025, all four major LHC experiments
have reported public machine-learning developments touching essentially every
stage of the data-processing chain: event generation and detector simulation
(Sect.~\ref{sec:sim}), the hardware and software trigger
(Sect.~\ref{sec:trigger}), and offline and online reconstruction
(Sect.~\ref{sec:reco}). Several of these methods are already in production, including the ALICE
generative models for TPC loopers, and the CMS and ATLAS anomaly-detection triggers (the first
unsupervised triggers operating in their respective experiments). Others, such as
the CMS Phase-2 Deep Sets jet tagger, LHCb GarNet, ATLAS GN3 flavour tagging
and MaskFormer tracking, and CMS MLPF, are proof-of-concept or R\&D efforts aimed at the
future HL-LHC. The consistent performance improvement pattern across
these examples illustrates how quickly machine
learning is maturing into a standard component of LHC data processing. The
ATLAS scaling study is one of the clearest indications of the potential of ML in high-energy physics:
current models are far from the limit of achievable performance, and the next
generation of algorithms will require both larger models and substantially
larger training datasets.

\printbibliography

\end{document}